\documentclass[11pt]{article}

\newsavebox{\foobox}
\newcommand{\slantbox}[2][0]{\mbox{%
        \sbox{\foobox}{#2}%
        \hskip\wd\foobox
        \pdfsave
        \pdfsetmatrix{1 0 #1 1}%
        \llap{\usebox{\foobox}}%
        \pdfrestore
}}
\newcommand\unslant[2][-.25]{\slantbox[#1]{$#2$}}

\newcommand{\mpi}{\text{\unslant[-.18]\pi}}
\newcommand{\mdelta}{\text{\unslant[-.18]\delta}}

\usepackage[left=2cm, right=2cm, top=2.5cm, bottom=2.5cm]{geometry}
\usepackage[x11names]{xcolor}
\usepackage{fancyhdr, amssymb, cancel, amsmath, graphicx, pgfplots, tikz}
\usepackage{isomath}

\usetikzlibrary{shadows}

\newcommand{\stylecolor}{blue!50!black}

\usepackage[labelfont={bf,sf, color=\stylecolor}, margin={1.5cm,0cm}]{caption}

\usepackage[colorlinks=true, urlcolor=\stylecolor!70!white, linkcolor=\stylecolor, citecolor=\stylecolor!70!white, hyperindex=true, linktocpage=true]{hyperref}

\usepackage[explicit]{titlesec}

\newcommand*\sectionlabel{}
\titleformat{\section}
  {\gdef\sectionlabel{}
   \Large\bfseries\scshape}
  {\gdef\sectionlabel{\thesection }}{0pt}
  {\begin{tikzpicture}[remember picture,overlay]
       \end{tikzpicture}
  }
\titlespacing*{\section}{0pt}{0pt}{0pt}

\newcommand*\subsectionlabel{}
\titleformat{\subsection}
  {\gdef\subsectionlabel{}
   \large\bfseries\scshape}
  {\gdef\subsectionlabel{\thesubsection  }}{0pt}
  {\begin{tikzpicture}[remember picture]
    	\draw (-0.15, 0) node[left] {\color{\stylecolor} \textsf{\subsectionlabel}};
	\draw (0.15, 0) node[right] {\color{\stylecolor} \textsf{#1}};
	\fill[color=\stylecolor] (-0.05, -0.23) rectangle (0.05, 0.23);
       \end{tikzpicture}
  }
\titlespacing*{\subsection}{-4pt}{10pt}{0pt}

\newcommand*\subsubsectionlabel{}
\titleformat{\subsubsection}
  {\gdef\subsubsectionlabel{}
   \bfseries\scshape}
  {\gdef\subsubsectionlabel{\thesubsubsection.\ \  }}{0pt}
  {\begin{tikzpicture}[remember picture]
    	\draw (0, 0) node[left] {\color{\stylecolor} \textsf{\subsubsectionlabel}};
	\draw (0, 0) node[right] {\color{\stylecolor} \textsf{#1}};
       \end{tikzpicture}
  }
\titlespacing*{\subsubsection}{-4pt}{7pt}{0pt}

\pgfplotsset{every axis legend/.append style={at={(1.02,1)},anchor=north west}}

\begin{document}

\allowdisplaybreaks

\pagestyle{fancy}
\renewcommand{\headrulewidth}{0pt}
\fancyhead{}

\fancyfoot{}
\fancyfoot[C] {\textsf{\textbf{\thepage}}}

\begin{equation*}
\begin{tikzpicture}
\draw (\textwidth, 0) node[text width = \textwidth, right] {\color{white} easter egg};
\end{tikzpicture}
\end{equation*}

\begin{equation*}
\begin{tikzpicture}
\draw (0.5\textwidth, -3) node[text width = \textwidth] {\huge  \textsf{\textbf{Kinetic theory of electronic transport in random \\ \vspace{0.07in} magnetic fields}} };
\end{tikzpicture}
\end{equation*}
\begin{equation*}
\begin{tikzpicture}
\draw (0.5\textwidth, 0.1) node[text width=\textwidth] {\large \color{black}  \textsf{Andrew Lucas}};
\draw (0.5\textwidth, -0.5) node[text width=\textwidth] {\small\textsf{Department of Physics, Stanford University, Stanford, CA 94305, USA}};
\end{tikzpicture}
\end{equation*}
\begin{equation*}
\begin{tikzpicture}
\draw (0, -13.1) node[right, text width=0.5\paperwidth] {\texttt{ajlucas@stanford.edu}};
\draw (\textwidth, -13.1) node[left] {\textsf{\today}};
\end{tikzpicture}
\end{equation*}
\begin{equation*}
\begin{tikzpicture}
\draw[very thick, color=\stylecolor] (0.0\textwidth, -5.75) -- (0.99\textwidth, -5.75);
\draw (0.12\textwidth, -6.25) node[left] {\color{\stylecolor}  \textsf{\textbf{Abstract:}}};
\draw (0.53\textwidth, -6) node[below, text width=0.8\textwidth, text justified] {\small  We present the theory of quasiparticle transport in perturbatively small inhomogeneous magnetic fields across the ballistic-to-hydrodynamic crossover.   In the hydrodynamic limit,  the resistivity $\rho$ generically grows proportionally to the rate of momentum-conserving electron-electron collisions at large enough temperatures $T$.   In particular, the resulting flow of electrons provides a simple scenario where viscous effects suppress conductance below the ballistic value.  This new mechanism for $\rho\propto T^2$ resistivity in a Fermi liquid may describe low $T$ transport in single-band $\mathrm{SrTiO}_3$.   };
\end{tikzpicture}
\end{equation*}

\tableofcontents

\begin{equation*}
\begin{tikzpicture}
\draw[very thick, color=\stylecolor] (0.0\textwidth, -5.75) -- (0.99\textwidth, -5.75);
\end{tikzpicture}
\end{equation*}

\titleformat{\section}
  {\gdef\sectionlabel{}
   \Large\bfseries\scshape}
  {\gdef\sectionlabel{\thesection }}{0pt}
  {\begin{tikzpicture}[remember picture]
	\draw (0.2, 0) node[right] {\color{\stylecolor} \textsf{#1}};
	\draw (0.0, 0) node[left, fill=\stylecolor,minimum height=0.27in, minimum width=0.27in] {\color{white} \textsf{\sectionlabel}};
       \end{tikzpicture}
  }
\titlespacing*{\section}{0pt}{20pt}{5pt}

\section{Introduction}
In a translation invariant metal, the electrical resistivity $\rho=0$ at finite density: adding some conserved momentum to the metal, we generate electric current that cannot decay.  But in a typical metal, $\rho >0$ because momentum is not conserved; $\rho$ then depends on how momentum relaxes.   Surprisingly,
 \begin{equation}
\rho \propto \frac{1}{\ell_{\mathrm{ee}}(T)}, \label{eq:main}
\end{equation}
is found experimentally in many correlated electron metals, including Fermi \cite{kadowaki, jacko} and non-Fermi \cite{mackenzie2013, hartnoll1} liquids; here $\ell_{\mathrm{ee}}$ is the mean free path for electron-electron collisions, and $T$ is temperature.  In a Fermi liquid,  $\ell_{\mathrm{ee}}\sim T^{-2}$.  There are two common mechanisms for (\ref{eq:main}):  (\emph{i})  Metals with  large Fermi  surfaces, where the Fermi wavelength of quasiparticles is comparable to the interatomic spacing, often have umklapp: momentum-relaxing electron-electron collisions \cite{ziman}.    (\emph{ii}) Baber found (\ref{eq:main}) in multi-band  Fermi liquids where current is carried by a light band, while a heavy band relaxes momentum  efficiently \cite{baber}.   In many metals including some transition metals and their oxides, heavy fermion  metals, and organic charge transfer  salts, (\emph{i}) or  (\emph{ii})  are consistent with experiment \cite{jacko}.

Recently, a curious observation of $\rho \propto T^2$ was reported in $\mathrm{SrTiO}_3$ \cite{behnia, stemmer}, at electronic carrier densities of $n \approx 10^{17}\; \mathrm{cm}^{-3}$.   A second band appears in $\mathrm{SrTiO}_3$ when $n\gtrsim 10^{18}  \; \mathrm{cm}^{-3}$; umklapp is allowed when $n\gtrsim 10^{20}\; \mathrm{cm}^{-3}$ \cite{behnia}.   We conclude that neither umklapp or  Baber scattering can explain the resistivity of low density $\mathrm{SrTiO}_3$: perhaps other routes to  (\ref{eq:main}) exist.

The simplest transport theory is the Drude model, where $\rho$ is proportional to the rate at which a single electron changes its direction of motion.    Umklapp and Baber scattering can both be accounted for in this framework.  However, this approach is inadequate if momentum-conserving electron-electron interactions are strong enough that the electrons flow collectively as a fluid \cite{gurzhi, spivak02, andreev, lucas, polini, levitovhydro, lucas3, alekseev, levitov1607, lucas1612, levitov1612, hartnoll1704, hartnoll1706}; see \cite{lucasreview2} for a review.   Experiments have found evidence of this hydrodynamic flow in some correlated metals \cite{molenkamp, bandurin, crossno, mackenzie, felser}.  However, a key prediction of hydrodynamics is that in a simple Fermi liquid, $\rho \propto \eta$, the shear viscosity, and $\eta \propto T^{-2}$ (up to logarithms \cite{novikov}).  The scaling $\rho \propto T^{-2}$ is not clearly seen in bulk resistivity measurements, however; evidence is strongest in flows through narrow  constrictions \cite{levitov1703}, and has also been found in magnetoresistance \cite{alekseev, kwwest}.  

 
 In this letter, we provide a simple hydrodynamic route to (\ref{eq:main}).  Under rather general circumstances,  (\ref{eq:main}) occurs when a viscous fluid flows through inhomogeneous magnetic fields.   We show this first using the Navier-Stokes equations, and then by solving the Boltzmann equation.  The latter approach allows us to further quantify the strength of hydrodynamic effects even when electron-electron scattering is weak.    As many correlated metals, including heavy fermions \cite{doniach, aynajian}, iron-based superconductors \cite{ypwu}, cuprates \cite{alloul, mendels, balatsky} and $\mathrm{SrTiO}_3$ \cite{brinkman, menyoung, moler} have magnetic disorder, our theory provides a simple way to reconcile (\ref{eq:main}) with hydrodynamics in unconventional metals --  including single-band $\mathrm{SrTiO}_3$ at low density.

\section{Viscous Transport}
We begin by computing the electrical  resistivity of a metal in which electrons flow as a  classical fluid, described by the linearized Navier-Stokes equations:
\begin{subequations}\label{eq:NS0}\begin{align}
0 &= \partial_i (nv_i) \\
-en B_{ij} v_j &= -en (E_i - \partial_i \mu) - \eta \partial_j \partial_j v_i.  \label{eq:NS}
\end{align}\end{subequations}
   $B_{ij} = \epsilon_{ijk}B_k$ represents the magnetic field, with $\epsilon_{ijk}$ the Levi-Civita tensor.     In order to compute the resistivity $E_i = \rho_{ij}J_j^{\mathrm{avg}}$, we must compute the spatially averaged charge current $J^{\mathrm{avg}}_j = -en\int \mathrm{d}^2\mathbf{x} \; v_j$ in an infinitesimally small electric field $E_i$.  We have not kept track of nonlinear terms (\ref{eq:NS0}); they contribute only to the nonlinear conductivity.    Hydrodynamics is valid when the  rate of momentum-conserving electron-electron  collisions, $\gamma \sim T^2/E_{\mathrm{F}}$, is faster than  other scattering rates (such as electron-impurity/phonon); here $E_{\mathrm{F}}$ is the Fermi energy.   This theory is naturally applicable to low density, low $E_{\mathrm{F}}$ metals  such as $\mathrm{SrTiO}_3$ or graphene \cite{bandurin, crossno, levitov1703}.

  Now, let us assume that the magnetic field $B_{ij}(\mathbf{x})$ is small, and that there is no uniform  background: $\int \mathrm{d}^d \mathbf{x} \; B_{ij} = 0$.    We look for a solution to (\ref{eq:NS0}) perturbatively in the strength of the magnetic field, which we denote $\delta$.   Letting $v_i(\mathbf{k}) \equiv \int  \mathrm{d}^d\mathbf{x}\; \mathrm{e}^{-\mathrm{i}\mathbf{k}\cdot\mathbf{x}} v_i(\mathbf{x})$, etc.,    we look for a  solution to (\ref{eq:NS0}) of the form $v = v_0 + v_1(\mathbf{x}) + \cdots$,  $\mu = \mu_1(\mathbf{x}) + \cdots$, with $v_0 \sim \delta^{-2}$, $v_1 \sim \delta^{-1}$, $\ldots$   \cite{andreev, lucas, lucas3}.   Most of the current arises at $\mathrm{O}(\delta^{-2})$, and is spatially homogeneous;  it is driven by the near-conservation of momentum.  Accordingly, we anticipate that $\rho \propto \delta^2$.     At $\mathrm{O}(\delta^{-1})$, (\ref{eq:NS0}) gives \begin{equation}
v_{1i}(\mathbf{k}) =-\frac{en}{\eta k^2} \left(\mdelta_{il} - \frac{k_ik_l}{k^2}\right)B_{lj}(\mathbf{k})v_{0j}. 
\end{equation}
Here $\mdelta_{ij}$  is the Kronecker delta function.
Averaging (\ref{eq:NS}) over space at $\mathrm{O}(\delta^0)$, we obtain \begin{equation}
E_i = \frac{1}{(2\mpi)^2} \int  \mathrm{d}^2\mathbf{k}  \; B_{ij}(-\mathbf{k}) v_{1j}(\mathbf{k}).
\end{equation}
Since $J_i  \approx -en v_{0i}$, the resistivity is
 \begin{equation}
\rho_{ij} = \frac{1}{\eta} \int \frac{\mathrm{d}^d\mathbf{k}}{(2\mpi)^d} B_{ik}(-\mathbf{k})\left(\frac{\mdelta_{kl}}{k^2} - \frac{k_kk_l}{k^4}\right) B_{lj}(\mathbf{k}).  \label{eq:visc}
\end{equation}
As $\eta \propto \ell_{\mathrm{ee}}$, we find (\ref{eq:main}).  

\section{Do Interactions Increase or Decrease the Resistivity?}
We have found a simple hydrodynamic mechanism for $\rho \propto 1/\ell_{\mathrm{ee}}$. Yet previous works \cite{gurzhi, spivak02, levitov1607, lucas1612, levitov1612} found that viscous flows lead to $\rho \propto \ell_{\mathrm{ee}}$.  Generalizing \cite{hartnoll1706}, we now explain this difference.

 
 Consider trying to push an electric current through an  inhomogeneous  landscape where $n$ varies from point to point.   In the absence of inhomogeneity, a current can flow without dissipation via a  uniform velocity field $v_{0i}$, and the system is globally in thermal equilibrium (at  finite momentum density).   Due to the inhomogeneity in $n$, $v_i$ cannot  be uniform: the charge current is $J_i=-env_i$, and its conservation requires $\partial_i J_i = 0$.  By simply modifying the velocity to $v_{0i} + v_{1i}(\mathbf{x})$, we recover conservation of charge, while maintaining local  thermal equilibrium.    $\rho$ is proportional to  the rate of momentum loss, $Dk_*^2$ (here $k_*$ is the typical wave number of the disorder, and $D\propto \eta$ is the momentum diffusion constant).  Thus we find $\rho \propto\ell_{\mathrm{ee}}$.

In contrast, as we push a uniform velocity through an inhomogeneous magnetic field, local stress $T_{ij}^{\mathrm{ext}}$ is created.   $T_{ij}^{\mathrm{ext}}$ must be balanced by internal fluid stress.   By changing the local density, we can maintain local thermal equilibrium while creating pressure gradients to cancel longitudinal external stresses.   However, pressure gradients do not cancel the external \emph{shear stress}.   These shear stresses are only cancelled by velocity \emph{gradients}, whose magnitudes scale inversely to  the viscosity:  $\partial v \sim T^{\mathrm{ext}}_{xy}/\eta$.   The amplitude of perturbations to the uniform velocity, $v_{1i}$, is enhanced by a factor $1/\ell_{\mathrm{ee}}k_*$.    Since the dissipative resistivity is proportional to the amplitude  of the perturbation squared,  we conclude that $\rho \propto (D k_*^2)/(\ell_{\mathrm{ee}}k_*)^2$.  Using $D\propto \eta \propto \ell_{\mathrm{ee}}$, we again obtain (\ref{eq:main}).

The key difference between potential and magnetic disorder is that the magnetic disorder creates flux of conserved quantities that cannot be compensated by changes in the local equilibrium parameters:  chemical potential $\mu$ and velocity $v_i$.   The necessary flow of the conserved shear momentum must be generated diffusively, by a velocity  gradient.   More generally, disorder that couples to  diffusive hydrodynamic modes typically leads to $\rho \propto 1/\ell_{\mathrm{ee}}$; indeed, diffusion signals the impossibility of transport in ideal hydrodynamics.  In contrast, disorder that couples to ballistic (sound) modes typically leads to $\rho \propto \ell_{\mathrm{ee}}$;  in this case, relevant conserved quantities can be transported while maintaining local equilibrium.  A formal derivation of these statements, and their generalizations to more exotic hydrodynamic models, are presented in Appendix \ref{app:rigor}.

\section{Boltzmann Transport}
Next, we show that under rather general circumstances,  explicit microscopic models exhibit the above viscous flows through inhomogeneous magnetic fields.  We consider a theory of weakly interacting fermionic quasiparticles of dispersion relation $\epsilon(\mathbf{p})$, in general spatial dimension $d$.  For simplicity, we neglect electron-phonon coupling and assume that the chemical potential $\mu$ can be chosen such that (\emph{i}) umklapp is negligible, and (\emph{ii}) $\mu \gg k_{\mathrm{B}}T$, so that thermal fluctuations about the Fermi surface are negligible.   Henceforth, we will take $\hbar=k_{\mathrm{B}}=1$ for simplicity.
On long length scales compared to the Fermi wavelength $\lambda_{\mathrm{F}}$, the dynamics of these quasiparticles is well described by (quantum) kinetic theory \cite{kamenev}.   
One can then use the (semi)classical Boltzmann equation to compute the resistivity, following \cite{hartnoll1706}.  Just as in hydrodynamics, it suffices to solve linearized equations to compute linear response properties such as $\rho_{ij}$.
Denoting the distribution function  of kinetic theory as $f(\mathbf{x},\mathbf{p})$, we write   \begin{equation}
 f(\mathbf{x},\mathbf{p}) \approx f_{\mathrm{eq}}(\mathbf{x},\mathbf{p}) - \frac{\partial f_{\mathrm{eq}}}{\partial \epsilon} \Phi(\mathbf{x},\mathbf{p}).  \label{eq:fxp}
 \end{equation} with $\Phi$ infinitesimally small, and \begin{equation}
 f_{\mathrm{eq}}(\mathbf{x},\mathbf{p}) \approx \frac{1}{1+\mathrm{e}^{(\epsilon(\mathbf{p})-\mu)/T}}.
 \end{equation}
It is useful to write  the $\mathbf{p}$-dependence of $\Phi$ in  Dirac bra-ket notation:
\begin{equation}
|\Phi(\mathbf{x})\rangle = \int \mathrm{d}^d\mathbf{p} \; \Phi(\mathbf{x},\mathbf{p}) |\mathbf{p}\rangle.
\end{equation}
The time-independent linearized Boltzmann equation is \begin{equation}
 \mathsf{L} |\Phi\rangle  +  \mathsf{W}|\Phi\rangle  = E_i |\mathsf{J}_i\rangle,  \label{eq:linboltz}
 \end{equation} with $|\mathsf{J}^i\rangle = -e \int \frac{\partial \epsilon}{\partial p_i} |\mathbf{p}\rangle$ denoting the electric current, \begin{equation}
 \mathsf{L} =  \frac{\partial \epsilon}{\partial p_i} \cdot \frac{\partial}{\partial x_i}  -e \frac{\partial \epsilon}{\partial p_i} B_{ij} \frac{\partial}{\partial p_j} 
 \end{equation}
 denoting the single  particle streaming operator, and $\mathsf{W}$ denoting a linearized collision operator, encoding the effects of momentum-conserving electron-electron collisions.  $\mathsf{W}$  is  symmetric and positive  semidefinite, and has null vectors associated with conservation of charge ($\Phi = 1$) and momentum ($\Phi =  p_i$).    
 We define the inner product \begin{equation}
\langle g_\alpha | g_\beta\rangle \equiv \int \frac{\mathrm{d}^d\mathbf{p}}{(2\mpi)^d} \left(-\frac{\partial f_{\mathrm{eq}}}{\partial \epsilon}\right) g_\alpha(\mathbf{p})g_\beta(\mathbf{p}).  \label{eq:innerproduct}
 \end{equation}
 
This  kinetic limit justifies neglecting the coupling of random magnetic fields to spin.   These effects contribute to (\ref{eq:linboltz}), and hence (\ref{eq:AJJ}), at subleading order in $\hbar$.

Following \cite{hartnoll1706}, we generalize  the derivation of (\ref{eq:visc}) to perturbatively solve (\ref{eq:linboltz}), assuming that $B_{ij}$  is perturbatively small and has zero average. Details are given in Appendix \ref{app:kinder}.  We find 
\begin{equation}
\rho_{ij} = \frac{1}{e^2n^2} \int \frac{\mathrm{d}^d\mathbf{k}}{(2\mpi)^d} B_{im}(-\mathbf{k})B_{jn}(\mathbf{k}) \mathcal{A}_{mn}(\mathbf{k}),   \label{eq:GRkin}
\end{equation}
where $n$ is the electron number density and \begin{equation}
\mathcal{A}_{mn}(\mathbf{k}) =  \langle \mathsf{J}_m| (\mathsf{W}+\mathsf{L}(\mathbf{k}))^{-1}|\mathsf{J}_n\rangle. \label{eq:AJJ}
\end{equation}
with  $\mathsf{L}(\mathbf{k})|\Phi\rangle \equiv  \int \mathrm{d}^d\mathbf{p} \; \mathrm{i}k_i\frac{\partial \epsilon}{\partial p_i} \Phi(\mathbf{p})|\mathbf{p}\rangle$.   
 (\ref{eq:GRkin}) also follows from the memory function formalism \cite{hofman, lucasMM, lucasreview}.  
 
 In order to explicitly evaluate (\ref{eq:GRkin}), we need to know $B_{ij}(\mathbf{k})$ and $\mathcal{A}_{ij}(\mathbf{k})$.  We will first describe $B_{ij}$, and then turn to $\mathcal{A}_{ij}$.    For a two dimensional Fermi liquid, a simple source of disorder is a density of $n_{\mathrm{imp}}$ magnetic dipoles of strength $\mathfrak{m}$ per unit area, oriented normal to and placed a distance $\xi$ above the electronic plane.  We assume that their locations above the electron liquid are random.   Averaging over the dipole positions, we show in Appendix \ref{app:dipoles} that \begin{equation}
\left\langle B_{ij}(-\mathbf{k})B_{kl}(\mathbf{k})\right\rangle =  n_{\mathrm{imp}}\left(\frac{\mu_0\mathfrak{m}}{2}\right)^2 \mathrm{e}^{-2k\xi} k^2  \epsilon_{ij}\epsilon_{kl} \label{eq:magdis}
\end{equation}
where $k=|\mathbf{k}|$, and $\epsilon_{ij}$ is the two-dimensional Levi-Civita tensor.   Schematically, we observe that $\rho_{ij}$ will be dominated by the value of $\mathcal{A}_{ij}$ when $k\sim 1/\xi$.   

\section{Circular Fermi Surfaces}
We now turn to the computation of $\mathcal{A}_{ij}$.  A solvable model where we can analytically compute $\mathcal{A}_{ij}$ is a two dimensional Fermi liquid with a circular Fermi surface of Fermi velocity $v_{\mathrm{F}}$ \cite{levitov1607}.   In the zero temperature limit, the Fermi function $f_{\mathrm{eq}}$ becomes a step function: $f_{\mathrm{eq}} \approx \mathrm{\Theta}(\mu-\epsilon)$.   In (\ref{eq:fxp}),  $\Phi$ is then multiplied by a delta function which enforces $|\mathbf{p}|=p_{\mathrm{F}}$.   Letting $\theta$ denote an angle on this circle ($\tan\theta = p_y/p_x$), we write
 \begin{equation}
\Phi = \sum_{n\in\mathbb{Z}} a_n(\mathbf{x}) \mathrm{e}^{\mathrm{i}n\theta}, \text{ or } |\Phi\rangle = \sum_{n\in\mathbb{Z}} a_n(\mathbf{x})|n\rangle.  \label{eq:Phian}
\end{equation}
Using (\ref{eq:innerproduct}), we find $\langle m|n\rangle = \nu \mdelta_{mn}$, where $\nu$ is the density of states at the Fermi surface and $\mdelta_{mn}$ is the Kronecker delta.   Using that $\mathsf{L}(\mathbf{k}) = \mathrm{i}k_x v_{\mathrm{F}}\cos\theta + \mathrm{i}k_yv_{\mathrm{F}}\sin\theta$, we explicitly compute the coefficients of the streaming matrix:
\begin{equation}
\mathsf{L}|n\rangle = \frac{v_{\mathrm{F}}}{2 } (\mathrm{i}k_x- k_y) |n+1\rangle + \frac{v_{\mathrm{F}}}{2 } (\mathrm{i}k_x+ k_y) |n-1\rangle .
\end{equation}
Assuming electron-electron interactions are rotationally symmetric, the collision operator is $\mathsf{W}|n\rangle = \gamma_n |n\rangle$. Charge and momentum conservation enforce $\gamma_{-1}=\gamma_0=\gamma_1=0$.

\begin{figure}[t]
\centering
\includegraphics{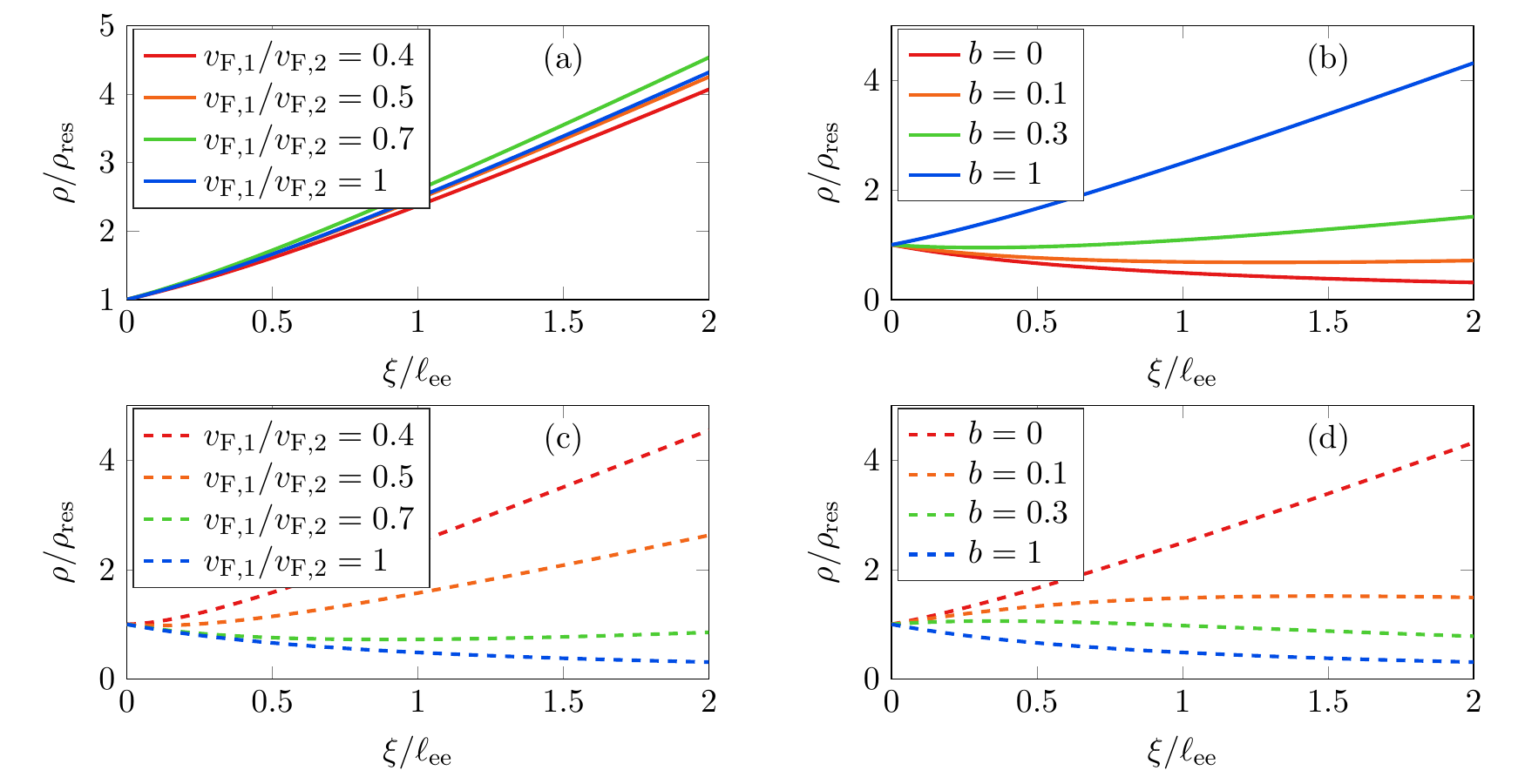}
\caption{(a) Resistivity in the two Fermi surface model with magnetic disorder;  (b) in the model (\ref{eq:pom}) with magnetic disorder;  (c) in the two Fermi surface model with  potential disorder;  (d) in  the model (\ref{eq:pom}) with potential disorder.   Data in (c,d) is taken from \cite{hartnoll1706}.   Resistivities are normalized by $\rho_{\mathrm{res}}$, the ballistic residual resistivity at $\ell_{\mathrm{ee}}=\infty$.  (\ref{eq:A1FS}) leads to the blue ($v_{\mathrm{F},1}/v_{\mathrm{F,2}}  =  1$) curve in panels (a,b).    We observe how whether electron-electron interactions enhance or suppress transport beyond the ballistic limit is sensitive to the particular hydrodynamic limit \emph{and} the nature of disorder.}
\label{fig:res}
\end{figure}

For certain choices of $\gamma_n$ for $|n|\ge 2$, we can analytically compute (\ref{eq:GRkin}).  An explicit example corresponds to the ``relaxation time" approximation \cite{bgk}: \begin{equation}
\gamma_n = \frac{v_{\mathrm{F}}}{\ell_{\mathrm{ee}}}, \;\;\; (|n|>1)  \label{eq:bgk}
\end{equation}
where $\ell_{\mathrm{ee}}$ is the mean free path of momentum-conserving collisions.  While it is unlikely that the collision integral takes this  form in a 2d Fermi liquid \cite{ledwith1, ledwith2}, this model correctly reproduces both ballistic and hydrodynamic limits.   $\mathcal{A}_{ij}$ has already been computed in this model \cite{levitov1607, lucas1612, hartnoll1706}; the result is
\begin{equation}
\mathcal{A}_{ij}(k) = e^2 v_{\mathrm{F}}\nu \frac{1+\sqrt{1+k^2\ell_{\mathrm{ee}}^2}}{k^2\ell_{\mathrm{ee}}} \left(\mdelta_{ij} - \frac{k_ik_j}{k^2}\right). \label{eq:A1FS}
\end{equation}
To derive this result, one uses the fact that $|\mathsf{J}_x\rangle \pm \mathrm{i}|\mathsf{J}_y\rangle = -ev_{\mathrm{F}}|\pm 1 \rangle$.
Combining (\ref{eq:GRkin}), (\ref{eq:magdis}) and (\ref{eq:A1FS}), we obtain $\rho$;  the result is shown in Figure \ref{fig:res}.

Let us unpack (\ref{eq:A1FS}).  In the ballistic limit $\ell_{\mathrm{ee}} \gg \xi$,  $\mathcal{A}_{ij} \propto 1/k $ is essentially independent of $\ell_{\mathrm{ee}}$, as is the resistivity $\rho$.  In this limit $\rho$ is the zero temperature residual resistivity:  transport is dominated by single-particle motion through the random magnetic fields.    The opposite limit $\ell_{\mathrm{ee}} \ll \xi$ is characterized by hydrodynamics, and so we should recover (\ref{eq:visc}).  We immediately observe that $\mathcal{A}_{ij} \propto 1/\ell_{\mathrm{ee}}$, consistent with (\ref{eq:visc}).   More quantitatively,  in the model (\ref{eq:bgk}), one finds the shear viscosity  \cite{lucas1612, hartnoll1706} \begin{equation}
\eta = \frac{n^2 \ell_{\mathrm{ee}}}{2\nu v_{\mathrm{F}}}.   \label{eq:visc2}
\end{equation}
Combining (\ref{eq:GRkin}), (\ref{eq:A1FS}) and (\ref{eq:visc2}), we indeed recover (\ref{eq:visc}).    Our microscopic model also allows us to quantitatively characterize the breakdown of  hydrodynamic transport when $\ell_{\mathrm{ee}}\sim\xi$.   Figure \ref{fig:res} demonstrates that even when $\ell_{\mathrm{ee}}=2\xi$, hydrodynamic effects can already double the resistivity above its residual value.   Hydrodynamic effects are  observable even when the condition $\ell_{\mathrm{ee}} \ll \xi$ fails.

The kinetic approach also allows us to study Fermi liquids with more complicated Fermi surfaces.  In a Fermi liquid with two small Fermi surfaces of Fermi velocities $v_{\mathrm{F,1}}$ and $v_{\mathrm{F,2}}$, the quasiparticle density on each pocket are both approximately conserved \cite{hartnoll1704, hartnoll1706}.    With this extra conservation law, the Navier-Stokes equations (\ref{eq:NS0}) are not applicable \cite{hartnoll1706}, and a more complicated (approximate) hydrodynamics emerges.  Following our heuristic discussion about when $\rho \propto \ell_{\mathrm{ee}}$ vs. $\rho \propto 1/\ell_{\mathrm{ee}}$, we expect that (\ref{eq:main}) continues to hold in the hydrodynamic limit, because shear momentum is still diffusive.     An explicit numerical computation in a toy model of this two-pocket Fermi liquid confirms these expectations: see Figure \ref{fig:res}.    Details of this toy model, which generalizes the single pocket model described explicitly above, are provided in Appendix \ref{app:pockets}.    We briefly note that for generic $v_{\mathrm{F,1}}/v_{\mathrm{F,2}}$, potential disorder also couples to a diffusive mode \cite{hartnoll1706}, and also leads to (\ref{eq:main}).   This effect is also shown in Figure \ref{fig:res}.

A final solvable model has \cite{hartnoll1706} 
\begin{equation}
\gamma_n = \frac{v_{\mathrm{F}}}{\ell_{\mathrm{ee}}} \times \left\lbrace \begin{array}{ll}  b &\  |n|=2 \\ 1 &\ |n|>2\end{array}\right..  \label{eq:pom}
\end{equation}
We will generally take $0<b<1$, so that quadrupolar deformations of the Fermi surface are long-lived excitations.    We find that (\ref{eq:A1FS}) generalizes to \begin{equation}
\mathcal{A}_{JJ}(k) = \frac{\nu}{v_{\mathrm{F}}} \frac{2b-1+\sqrt{1+k^2\ell_{\mathrm{ee}}^2}}{k^2\ell_{\mathrm{ee}}}.   \label{eq:AbFS}
\end{equation}

Figure \ref{fig:res} again shows $\rho$ as a function of $b$ and $\ell_{\mathrm{ee}}$ in this model.   The case $b=0$ is most interesting:  here we observe that potential disorder leads to $\rho\propto 1/\ell_{\mathrm{ee}}$ while magnetic disorder leads to $\rho \propto \ell_{\mathrm{ee}}$.  This unintuitive behavior is opposite of  (\ref{eq:bgk}).    To explain how this behavior follows from the conservation of quadrupolar deformations of the Fermi surface, we compute the hydrodynamic normal modes of this model in Appendix  \ref{app:b0}.       What  we find is that the hydrodynamics of this  model is rather different from (\ref{eq:NS0}).    For wave number $\mathbf{k}=k\hat{\mathbf{x}}$, we observe two ballistically propagating modes:  one coupling deformations of the Fermi surface of $p_y$ and $d_{xy}$ types, and another coupling $s$, $p_x$ and $d_{x^2-y^2}$ types.  The lone diffusive mode now corresponds to out-of-phase $s$ and $d_{x^2-y^2}$ deformations.   Because current/momentum couple only to ballistic modes, magnetic disorder leads to $\rho \propto \ell_{\mathrm{ee}}$; as density now couples to a diffusive mode, we find $\rho \propto 1/\ell_{\mathrm{ee}}$ with potential disorder.


In typical Fermi liquids,  transverse momentum is a diffusive quantity:  quadrupolar  fluctuations of the Fermi surface are not typically conserved.    As we detail in Appendix \ref{app:rigor}, this implies that the interplay of viscous hydrodynamic and magnetic disorder will typically lead to (\ref{eq:main}), providing another route to $T^2$ resistivity in Fermi liquids.

\section{Application to $\boldsymbol{\mathrm{SrTiO}_3}$}

In this letter  we have shown that (\ref{eq:main}) can result from the interplay of electron-electron interactions and the magnetic impurities.  We now argue why this theory may explain the puzzling resistivity of low density $\mathrm{SrTiO}_3$.   (\emph{i}) Magnetic disorder in $\mathrm{SrTiO}_3$ is well-established  \cite{brinkman, menyoung, moler}.  (\emph{ii}) Letting $\rho \approx \rho_0 + AT^2$,  \cite{behnia} found that $A$ is insensitive to the emergence of a second band of carriers at $n\sim 10^{18} \; \mathrm{cm}^{-3}$, as is our toy model of a multi-band metal in Figure \ref{fig:res}a.   (\emph{iii})  The residual resistivity $\rho_0$ appears unrelated to $A$ \cite{behnia, stemmer}.  In the weak disorder limit, contributions to $\rho(T)$ from magnetic vs. potential disorder add via Matthiesen's rule.  Potential disorder  affects $\rho$ most strongly at low temperatures: see Figure \ref{fig:res}c.   We speculate  that $\rho_{T\rightarrow 0}$ is dominated by potential disorder, while $A$ is dominated by magnetic disorder.   We predict that changing the density and/or strength of magnetic/potential impurities would most strongly affect the measured coefficients of $T^2/T^0$ in $\rho(T)$, respectively.   
  (\emph{iv}) $A$ has been found to be an unconventional function of carrier density $n$ \cite{stemmer, behnia}.   Using (\ref{eq:visc}) and (\ref{eq:visc2}), our hydrodynamic mechanism for $T^2$ resistivity suggests that $A \propto n_{\mathrm{imp}} \nu/ n^2 E_{\mathrm{F}}$.   We propose that correlations between the amount of oxygen depletion, magnetic inhomogeneity and $n$ are responsible for the unconventional $A(n)$ observed in $\mathrm{SrTiO}_3$.
       
The resistivity of $\mathrm{SrTiO}_3$ can also be measured to temperature $T\gtrsim T_{\mathrm{F}}$, where thermal diffusion will also become important.   A microscopic computation of the kinetic coefficients of $\mathrm{SrTiO}_3$ could determine the quantitative impact of this additional  mode on $\rho(T)$.

\addcontentsline{toc}{section}{Acknowledgements}
\section*{Acknowledgements}

I thank Kamran Behnia, Tomas Bzdu\v{s}ek, Sean Hartnoll, Connie Mousatov and Suzanne Stemmer for useful discussions.   I am supported by the Gordon and Betty Moore Foundation's EPiQS Initiative through Grant GBMF4302.

\begin{appendix}
\section{When Interactions Increase or Decrease the Resistivity}
\label{app:rigor}
In this appendix, we  use the kinetic formalism described in the main text to make the intuitive argument for when $\rho \propto \ell_{\mathrm{ee}}$ vs. $\rho \propto 1/\ell_{\mathrm{ee}}$ in the hydrodynamic regime rigorous.    We decompose the set of all $|\Phi(\mathbf{p})\rangle$ into three groups such that \begin{subequations}\begin{align}
\mathsf{W} &= \left( \begin{array}{ccc} 0 &\ 0 &\ 0 \\ 0 &\ 0 &\ 0 \\ 0 &\ 0 &\ \mathsf{W}_0 \end{array}\right), \\
\mathsf{L}(\mathbf{k}) &= \left( \begin{array}{ccc} \mathsf{L}_1 &\ 0 &\ -\mathsf{L}_2^\dagger\\ 0 &\ 0 &\ - \mathsf{L}_3^\dagger \\ \mathsf{L}_2 &\ \mathsf{L}_3 &\ \mathsf{L}_4 \end{array}\right),
\end{align}\end{subequations}
with $\mathsf{L}_1$ invertible.   The first row (I) of $|\Phi\rangle$ contains conserved quantities which mix among themselves under streaming (analogous to sound waves),  the second row (II) of $|\Phi\rangle$ contains conserved quantities which only couple via streaming to non-conserved modes (purely diffusive hydrodynamic modes), and the third row (III) of $|\Phi\rangle$ corresponds to non-conserved modes.     In the hydrodynamic limit $k\ell_{\mathrm{ee}} \ll 1$, using the scalings $\mathsf{L} \sim k$ and $\mathsf{W}_0 \sim \ell_{\mathrm{ee}}^{-1}$, along with block matrix inversion identities,  we find that for any vector $|\Phi\rangle$ which is entirely even or entirely odd under time reversal,
\begin{equation}
\langle \Phi| (\mathsf{W}+\mathsf{L})^{-1} |\Phi\rangle \lesssim \langle \Phi| \left( \begin{array}{ccc} \ell_{\mathrm{ee}} &\ k^{-1} &\ k^{-1} \\ k^{-1} &\ \ell_{\mathrm{ee}}^{-1}k^{-2} &\ k^{-1} \\ k^{-1} &\ k^{-1} &\ \ell_{\mathrm{ee}} \end{array}\right) |\Phi\rangle.  \label{eq:WL}
\end{equation}
$\mathcal{A}_{ij} \sim \ell_{\mathrm{ee}}^{-1} k^{-2}$ if and only if $|\mathsf{J}_i\rangle$ has non-vanishing weight in block row II.   In more physical terms, the current must have some overlap with a diffusive hydrodynamic mode, such as shear momentum.    

The inequality in (\ref{eq:WL}) -- at least for the block diagonal elements -- appears to be qualitatively saturated in all toy models that have been studied to date.   An informal (but still rather technical) proof that this saturation is generic (and will be challenging to violate in non-pathological models) is as follows.   

First, let us consider the top left block of (\ref{eq:WL}), which we denote as \begin{equation}
(\mathsf{W}+\mathsf{L})^{-1}_{\mathrm{I,I}} = \left(\mathsf{L}_1 + \mathsf{L}_2^\dagger (\mathsf{W}_0+\mathsf{L}_4)^{-1}\mathsf{L}_2\right)^{-1} \approx \left(\mathsf{L}_1 + \mathsf{L}_2^\dagger \mathsf{W}_0^{-1}\mathsf{L}_2\right)^{-1}.  \label{eq:IIinv}
\end{equation}  The approximate equality above is exact up to subleading corrections (componentwise) in $k\ell_{\mathrm{ee}}$.   Because $\mathsf{L}_1$ is the projection of the Fourier transform of an antisymmetric matrix -- $\frac{\partial \epsilon}{\partial p_i} \frac{\partial}{\partial x_i}$ projected onto  conserved modes -- its eigenvalues will come in pairs $\pm \mathrm{i}\lambda(\mathbf{k})$.    So let us consider a basis for block row I where \begin{equation}
\mathsf{L}_1 = \mathrm{i} \left(\begin{array}{ccc} \lambda_1 \sigma^x &\ 0 &\ 0 \\ 0 &\ \lambda_2 \sigma^x &\ \cdots \\ 0 &\ \vdots &\ \ddots \end{array}\right).
\end{equation}
As $\mathsf{L}_1$ is odd under time reversal, one of the basis vectors in each of the block rows above is odd under time reversal, and the other is even.   Because $\mathsf{L}_2^\dagger \mathsf{W}_0^{-1}\mathsf{L}_2$ is symmetric and positive-definite, if there is any vector $|\Phi_{\mathrm{I}}\rangle$ obeying $\langle \Phi_{\mathrm{I}}| \mathsf{L}_2^\dagger \mathsf{W}_0^{-1}\mathsf{L}_2|\Phi_{\mathrm{I}}\rangle = 0$, then $\langle \Phi_{\mathrm{I}}| \mathsf{L}_2^\dagger \mathsf{W}_0^{-1}\mathsf{L}_2|\Psi_{\mathrm{I}}\rangle = 0$ for any vector $|\Psi_{\mathrm{I}}\rangle$ in the row I subspace.   This identity implies that each eigenvector of $\mathsf{L}_1$ obeys either (\emph{i}) $\langle \Phi_{\mathrm{I}}| \mathsf{L}_2^\dagger \mathsf{W}_0^{-1}\mathsf{L}_2|\Phi_{\mathrm{I}}\rangle > 0$ or (\emph{ii}) $\langle \Phi_{\mathrm{I}}| (\mathsf{L}_1+\mathsf{L}_2^\dagger \mathsf{W}_0^{-1}\mathsf{L}_2)^{-1}|\Phi_{\mathrm{I}}\rangle =  \langle \Phi_{\mathrm{I}}| \mathsf{L}_1^{-1}|\Phi_{\mathrm{I}}\rangle$.   If $|\Phi_{\mathrm{I}}\rangle$ transforms with definite sign under time reversal, $\langle \Phi_{\mathrm{I}}| \mathsf{L}_1^{-1}|\Phi_{\mathrm{I}}\rangle=0$.   In general, we do not expect modes with exactly vanishing spectral weight to exist, barring exact identities such as Ward identities (see (\ref{eq:ward}) below) that demand it.   We conclude that under generic circumstances, $\langle \Phi_{\mathrm{I}}| \mathsf{L}_2^\dagger \mathsf{W}_0^{-1}\mathsf{L}_2|\Phi_{\mathrm{I}}\rangle > 0$.   As $\mathsf{L}_1 \sim k$ and $\mathsf{L}_2^\dagger \mathsf{W}_0^{-1}\mathsf{L}_2 \sim k^2$, in the limit $k\rightarrow 0$ we  estimate that for any $|\Phi_{\mathrm{I}}\rangle$ transforming with definite sign under time reversal, \begin{equation}
\langle \Phi_{\mathrm{I}}| \left(\mathsf{L}_1 + \mathsf{L}_2^\dagger \mathsf{W}_0^{-1}\mathsf{L}_2\right)^{-1} |\Phi_{\mathrm{I}}\rangle \sim \left(\begin{array}{cc} 1 &\ 0 \end{array}\right)\left(\begin{array}{cc} \ell_{\mathrm{ee}}k^2 &\  \mathrm{i}k \\ \mathrm{i}k &\ \ell_{\mathrm{ee}}k^2 \end{array}\right) \left(\begin{array}{c} 1 \\ 0 \end{array}\right) \sim \ell_{\mathrm{ee}}.
\end{equation}

The remaining two blocks are much simpler. Analogously to (\ref{eq:IIinv}), we find \begin{equation}
(\mathsf{W}+\mathsf{L})^{-1}_{\mathrm{II,II}} \approx \left(\mathsf{L}_3^\dagger \mathsf{W}_0^{-1}\mathsf{L}_3 \right)^{-1}.
\end{equation}   The assumption that there are no non-dynamical spatial inhomogeneities in kinetic theory implies, as before, that  $\mathsf{L}_3^\dagger \mathsf{W}_0^{-1}\mathsf{L}_3$ is non-singular.     Using the scalings above, we arrive at  $(\mathsf{W}+\mathsf{L})^{-1}_{\mathrm{II,II}} \sim \ell_{\mathrm{ee}}^{-1}k^{-2}$.  Finally, in block III,   $\mathsf{L}$ is negligible in the $k\rightarrow 0$ limit:  $(\mathsf{W}+\mathsf{L})^{-1}_{\mathrm{III,III}} \approx \mathsf{W}_0^{-1} \sim \ell_{\mathrm{ee}}$.

The scalings derived above of $(\mathsf{W}+\mathsf{L})^{-1}_{\mathrm{I,I}}$ and $(\mathsf{W}+\mathsf{L})^{-1}_{\mathrm{II,II}}$ are consistent with the hydrodynamic spectral weights associated with conserved quantities that either propagate ballistically or diffusively, respectively \cite{kadanoff, kovtun};  see also \cite{hartnoll1706}.   This explains our argument in the main text that we can check whether $\rho\propto \ell_{\mathrm{ee}}$ or $\rho \propto 1/\ell_{\mathrm{ee}}$ in the hydrodynamic limit by checking whether the impurities couple to sound waves or diffusion, respectively.

Returning to the focus of this paper on magnetic disorder, it remains to understand whether the current operator is generically contained within blocks I, II, and/or III.    The key observation is that at finite density, the current $|\mathsf{J}_i\rangle$ always overlaps with momentum $|\mathsf{P}_i\rangle$:  for a precise statement, see (\ref{eq:JP}) below.   Transverse momentum $|\mathsf{P}_i\rangle$, obeying $k_i|\mathsf{P}_i\rangle = 0$, often only couples to non-conserved modes, and is therefore in block II.   

In some model Fermi liquids, it may be possible to explicitly prove that transverse momentum is diffusive (in block II) using group theory.   As an example, suppose that all inversion-even conserved quantities $|\rho\rangle$ transform in the trivial representation $\mathbf{1}$ of the point group $G$ of the crystal, while $|\mathsf{P}_i\rangle$ and $|\mathsf{J}_i\rangle$ transform in represetation $\mathbf{V}$.         Note that $\epsilon(\mathbf{p})$, $f_{\mathrm{eq}}$ and the inner product (\ref{eq:innerproduct}) are invariant under the action of $G$ on $\mathbf{p}$, so the overlap of $|\mathsf{J}_i\rangle$ and $|\mathsf{P}_i\rangle$ implies they transform in the same representation (see Section 7.1 of \cite{dresselhaus}).   The same orthogonality theorems also imply  that $\mathbf{V} \otimes \mathbf{V} = \mathbf{1} \oplus \cdots$ contain \emph{exactly one} copy of the trivial representation $\mathbf{1}$, and that \begin{equation}
\langle \rho | \mathsf{L}(k_i)|\mathsf{P}_j\rangle \sim \mdelta_{ij},  \label{eq:LP}
\end{equation}
and inversion symmetry demands that $ \mathsf{L}(k_i)|\mathsf{P}_j\rangle$ has no overlap with any inversion odd basis vectors.  We conclude that the streaming terms $\mathsf{L}$ couple transverse momenta only to non-conserved inversion-even modes.  Conservation of  transverse momentum, together with (\ref{eq:WL}), implies (\ref{eq:main}).   

In the model (\ref{eq:pom}) with $b=0$, the conserved $|\pm 2\rangle$ modes are inversion even, and so the paragraph above does not apply.  However, the group theoretical arguments above suffice to show that the interplay of magnetic disorder and hydrodynamic effects lead to (\ref{eq:main}) in many anisotropic electron fluids.


\section{Resistivity from the Boltzmann Equation}\label{app:kinder}
In this appendix, we derive (\ref{eq:GRkin}) and (\ref{eq:AJJ}), following \cite{hartnoll1706}.   The logic follows the hydrodynamic derivation of (\ref{eq:visc}).   We begin by noting that when $B_{ij}=0$ and $E_i=0$, an exact time-independent solution to (\ref{eq:linboltz}) is \begin{equation}
|\Phi \rangle =  \alpha_i |\mathsf{P}_i\rangle,  \label{eq:phiisp}
\end{equation}
where $|\mathsf{P}_i\rangle = \int \mathrm{d}^d \mathbf{p} \; p_i|\mathbf{p}\rangle$, for any $\alpha$.    Our goal is now  to show that when $E_i\ne 0$ and $B_{ij}\ne 0$, then  we can perturbatively construct a solution to (\ref{eq:linboltz}), analogous to the Navier-Stokes equations, starting from (\ref{eq:phiisp}).   We write \begin{equation}
|\Phi\rangle = \alpha_i |\mathsf{P}_i\rangle + |\Phi_1(\mathbf{x})\rangle + \cdots 
\end{equation}
with $\alpha_i \sim \delta^{-2}$, $|\Phi_1\rangle \sim \delta^{-1}$, etc.   The Fourier  transform of (\ref{eq:linboltz}) at $\mathrm{O}(\delta^{-1})$ reads \begin{equation}
(\mathsf{W}+\mathsf{L}(\mathbf{k}))|\Phi_1(\mathbf{k})\rangle = e \frac{\partial \epsilon}{\partial p_i} B_{ij}(\mathbf{k}) \alpha_j|\mathsf{n}\rangle = - \alpha_j B_{ij}(\mathbf{k})|\mathsf{J}_i\rangle.
\end{equation}
where $|\mathsf{n}\rangle = \int \mathrm{d}^d\mathbf{p} \; |\mathbf{p}\rangle$ corresponds to the vector associated with density fluctuations in the distribution function.     At $\mathrm{O}(\delta^0)$, we spatially integrate (\ref{eq:linboltz}), sandwiched with the bra $\langle \mathsf{P}_k|$, to fix $\alpha_i$.   Using the identities \begin{subequations}\begin{align}
\langle \mathsf{J}_i|\mathsf{P}_j\rangle &= -e\int \frac{\mathrm{d}^d\mathbf{p}}{(2\mpi\hbar)^d} \left(-\frac{\partial f_{\mathrm{eq}}}{\partial \epsilon} \frac{\partial \epsilon}{\partial p_i}\right) p_ j =-e \int \frac{\mathrm{d}^d\mathbf{p}}{(2\mpi\hbar)^d} f_{\mathrm{eq}} \frac{\partial p_j}{\partial p_i} = -e n\mdelta_{ij},  \label{eq:JP} \\
\left\langle \mathsf{P}_k \left| \frac{\partial \epsilon}{\partial p_i} B_{ij}(\mathbf{k}) \frac{\partial}{\partial p_j} \right|\Phi\right\rangle &= \int \frac{\mathrm{d}^d\mathbf{p}}{(2\mpi\hbar)^d} f_{\mathrm{eq}} \frac{\partial}{\partial p_i} \left(B_{ij}(\mathbf{k}) p_k \frac{\partial \Phi}{\partial  p_j}\right) =  \int \frac{\mathrm{d}^d\mathbf{p}}{(2\mpi\hbar)^d} f_{\mathrm{eq}} B_{kj}(\mathbf{k})  \frac{\partial \Phi_2}{\partial  p_j} \notag \\
&= \int \frac{\mathrm{d}^d\mathbf{p}}{(2\mpi\hbar)^d} \left(-\frac{\partial f_{\mathrm{eq}}}{\partial \epsilon}\right) \frac{\partial \epsilon}{\partial p_j} B_{kj}(\mathbf{k}) \Phi_2 = -\frac{1}{e}\langle \mathsf{J}_j| B_{jk}(\mathbf{k}) |\Phi\rangle,
\end{align}\end{subequations}
we find
 \begin{equation}
\langle \mathsf{P}_k | E_i |\mathsf{J}_i\rangle = -enE_k =  \int \frac{\mathrm{d}^d\mathbf{k}}{(2\mpi)^d} B_{ij}(-\mathbf{k}) \langle \mathsf{J}_j |\Phi_1(\mathbf{k})\rangle = \alpha_k \int \frac{\mathrm{d}^d\mathbf{k}}{(2\mpi)^d} B_{ij} (-\mathbf{k})  B_{kl}(\mathbf{k})  \mathcal{A}_{jl}(\mathbf{k}).   \label{eq:appB1}
\end{equation} 
The spatially averaged current  is -- by definition -- given by $J^{\mathrm{avg}}_i = \langle \mathsf{J}_i| \Phi\rangle$.   To leading order, \begin{equation}
J^{\mathrm{avg}}_i =  \alpha_k \langle \mathsf{J}_i|\mathsf{P}_k\rangle = -en\alpha_i.  \label{eq:appB2}
\end{equation}
Combining (\ref{eq:appB1}) and (\ref{eq:appB2}), we  obtain (\ref{eq:AJJ}).

Using the Ward identity \begin{equation}
k_i \mathcal{A}_{ij}(\mathbf{k}) = k_j \mathcal{A}_{ij}(\mathbf{k}) = 0,  \label{eq:ward}
\end{equation}
we obtain the  result predicted  by the  memory matrix formalism \cite{hofman, lucasMM, lucasreview}: \begin{equation}
\rho_{ij} = \frac{1}{e^2n^2}\int \frac{\mathrm{d}^d\mathbf{k}}{(2\mpi)^d}  k_i k_j  a_k(-\mathbf{k}) a_l(\mathbf{k}) \mathcal{A}_{kl}(\mathbf{k}),  \label{eq:memm}
\end{equation}


\section{Magnetic Fields from Random Magnetic Dipoles}
\label{app:dipoles}
In this appendix we explicitly derive (\ref{eq:magdis}).    For a magnetic dipole of strength $\mathfrak{m} \hat{\mathbf{z}}$ located a distance $\xi$ above a two dimensional plane at the point $(x,y,z) = (0,0,\xi)$, the magnetic vector potential lies entirely within the $xy$ plane.  Tthe $x$ and $y$ components of this vector potential are given by \begin{equation}
A_i = \frac{\mu_0\mathfrak{m}}{4\mpi} \frac{-\epsilon_{ij}x_j}{(r^2+\xi^2)^{3/2}} = -\frac{\mu_0\mathfrak{m}}{4\mpi}\epsilon_{ij}\frac{\partial}{\partial x_j} \frac{1}{\sqrt{r^2+\xi^2}}.  \label{eq:Aix}
\end{equation}
Our first goal is to Fourier transform $A_i$ in the $xy$ plane.  It is easier to ``guess" the transform $A_i(\mathbf{k})$ and check that the Fourier transform of $A_i(\mathbf{k})$ is (\ref{eq:Aix}).   Our ``guess" for the Fourier transform is \begin{equation}
A_i(\mathbf{k}) = -\mathrm{i}\frac{\mu_0\mathfrak{m}}{2} \epsilon_{ij} \frac{k_j}{k}\mathrm{e}^{-k\xi},
\end{equation}
where we have denoted $k=|\mathbf{k}|$.    The Fourier transform of $A_i(\mathbf{k})$ can be analytically computed:  \begin{align}
A_i(\mathbf{x}) &= \int \frac{\mathrm{d}^2\mathbf{k}}{(2\mpi)^2} \left( -\mathrm{i}\frac{\mu_0\mathfrak{m}}{2} \epsilon_{ij} \frac{k_j}{k}\mathrm{e}^{-k\xi}\right) \mathrm{e}^{\mathrm{i}\mathbf{k}\cdot\mathbf{x}} = -\frac{\partial}{\partial x_j} \epsilon_{ij} \int \frac{\mathrm{d}^2\mathbf{k}}{(2\mpi)^2}\frac{\mu_0\mathfrak{m}}{2} \frac{\mathrm{e}^{-k\xi}}{k} \notag \\
&= -\frac{\partial}{\partial x_j} \epsilon_{ij} \frac{\mu_0\mathfrak{m}}{4\mpi}\int\limits_0^{2\mpi} \frac{\mathrm{d}\theta}{2\mpi} \int\limits_0^\infty \mathrm{d}k  \mathrm{e}^{\mathrm{i}kr\cos\theta-k\xi} = -\frac{\partial}{\partial x_j} \epsilon_{ij} \frac{\mu_0\mathfrak{m}}{4\mpi}\int\limits_0^{2\mpi} \frac{\mathrm{d}\theta}{2\mpi} \frac{1}{\xi - \mathrm{i}r\cos\theta}.
\end{align}
This final integral can be done via contour integration, and one indeed recovers (\ref{eq:Aix}). 
In two dimensions, the Fourier transform of the magnetic field $B_{ij} = \mathrm{i}k_i A_j - \mathrm{i}k_j A_i \equiv B \epsilon_{ij}$, where \begin{equation}
B = \mathrm{i} \epsilon_{ij} k_i A_j = \frac{\mu_0\mathfrak{m}}{2} k \mathrm{e}^{-k\xi}.
\end{equation}

If, instead of being located above the origin, the magnetic dipole is located at the point $(x,y,\xi)$, then we simplify multiply $A_i(\mathbf{k})$ by a factor of $\mathbf{e}^{\mathrm{i}(k_xx+k_yy)}$.    If we have multiple magnetic dipoles located at points $\mathbf{r}_a$, then \begin{equation}
A_i(\mathbf{k}) = \mathrm{i}\frac{\mu_0\mathfrak{m}}{2} \epsilon_{ij} \frac{k_j}{k}\mathrm{e}^{-k\xi} \sum_a \mathrm{e}^{\mathrm{i}\mathbf{k}\cdot \mathbf{r}_a}. 
\end{equation}
So before disorder averaging, the magnetic fields in (\ref{eq:GRkin}) take the form \begin{equation}
B_{ij}(-\mathbf{k}) B_{kl}(\mathbf{k}) = \epsilon_{ij}\epsilon_{kl} \left(\frac{\mu_0\mathfrak{m}}{2}\right)^2 k^2 \mathrm{e}^{-2k\xi} \sum_{a,b} \mathrm{e}^{\mathrm{i}\mathbf{k}\cdot( \mathbf{r}_a - \mathbf{r}_b)}.
\end{equation}
Disorder averaging corresponds to spatially averaging over all possible positions of $\mathbf{r}_a$ for each $a$: \begin{equation}
\langle F\rangle_{\mathrm{dis}} = \int \left( \prod_a \frac{\mathrm{d}^2\mathbf{r}_a}{V}\right) F.
\end{equation}
where $V$ is the volume of the  sample.   The only non-vanishing terms in $\langle  B_{ij}(-\mathbf{k}) B_{kl}(\mathbf{k}) \rangle_{\mathrm{dis}}$ correspond to $a=b$.   If there are $N_{\mathrm{imp}} = n_{\mathrm{imp}}V$ total impurities in the volume $V$, we conclude that  \begin{equation}
\langle  B_{ij}(-\mathbf{k}) B_{kl}(\mathbf{k}) \rangle_{\mathrm{dis}} = \epsilon_{ij}\epsilon_{kl} \left(\frac{\mu_0\mathfrak{m}}{2}\right)^2 k^2 \mathrm{e}^{-2k\xi}  \times \frac{N_{\mathrm{imp}}}{V}
\end{equation}
which is equivalent to (\ref{eq:magdis}).

\section{A Model of  Two Fermi Surfaces}
\label{app:pockets}
We review the model of two Fermi surfaces described in \cite{hartnoll1706}.    For simplicity, we assume that the density of states of each Fermi surface is the  same, $\nu$.  We also assume that each band has a quadratic dispersion relation with the same mass $m$, so that the current and momentum of each band separately are proportional to each other.   If there are two circular Fermi surfaces, then we may generalize the logic of the main text to replace (\ref{eq:Phian}) with \begin{equation}
|\Phi\rangle = \sum_{A=1}^2 \sum_{n\in\mathbb{Z}} a_{nA} |nA\rangle .
\end{equation}
in the low temperature  limit.   The index $A=1,2$ denotes which one of the two Fermi surfaces carries a fluctuation.    The streaming matrix  is \begin{equation}
\mathsf{L}|nA\rangle = \frac{v_{\mathrm{F}}}{2 } (\mathrm{i}k_x- k_y) |(n+1)A\rangle + \frac{v_{\mathrm{F}}}{2 } (\mathrm{i}k_x+ k_y) |(n-1)A\rangle,  \label{eq:LappD}
\end{equation}
and the collision matrix is \begin{equation}
\mathsf{W} = \frac{v_{\mathrm{F2}}}{\nu \ell_{\mathrm{ee}}}\left[\sum_{A,|j|\ge 2} |jA\rangle\langle jA| + \sum_{j=\pm 1} \frac{(v_{\mathrm{F2}}|j1\rangle -v_{\mathrm{F1}}|j2\rangle )(v_{\mathrm{F2}}\langle j1| -v_{\mathrm{F1}}\langle j2| ) }{v_{\mathrm{F1}}^2+v_{\mathrm{F2}}^2}\right]  \label{eq:WappD}
\end{equation}
The factor of $\nu$ is required due to the non-trivial inner product (\ref{eq:innerproduct}).

In this model, the total momentum is simply given by the momentum carried in each band.   With the dispersion relation described above, we find that the total momentum and current are proportional, and given by \cite{hartnoll1706} \begin{equation}
|\mathsf{J}_x\rangle \pm \mathrm{i}|\mathsf{J}_y\rangle \equiv |\mathsf{J}_\pm \rangle = -ev_{\mathrm{F,1}} |\pm 1,1\rangle -ev_{\mathrm{F,2}} |\pm 1,2\rangle.
\end{equation}
The techniques to compute $\langle \mathsf{J}_i| (\mathsf{W}+\mathsf{L})^{-1}|\mathsf{J}_j\rangle$ are detailed in \cite{levitov1607, lucas1612, hartnoll1706}.
They amount to a  more souped up version of the following argument, which is sufficient to understand the hydrodynamic limit.   Consider the block matrix inversion identity  \begin{equation}
\left(\begin{array}{cc}  \mathbf{v}^{\mathsf{T}} &\ 0 \end{array}\right)\left(\begin{array}{cc}  \mathsf{A} &\ \mathsf{B} \\ \mathsf{C} &\ \mathsf{D} \end{array}\right)^{-1} \left(\begin{array}{c}  \mathbf{w} \\ 0 \end{array}\right) = \mathbf{v}^{\mathsf{T}} \left(\mathsf{A} - \mathsf{BD}^{-1}\mathsf{C} \right)^{-1}\mathbf{w}.   \label{eq:blockmatrix}
\end{equation}
We arrange the infinite dimensional vector space spanned by $|nA\rangle$ such that the modes $|jA\rangle$ with $|j| \le 1$ are in the top block, and modes with $|j|>1$ are in the bottom block.    Using (\ref{eq:LappD}) and (\ref{eq:WappD}), we see that to leading order as $k \rightarrow 0$, the inverse of $\mathsf{W}+\mathsf{L}$ in the bottom block is simply $\mathsf{W}^{-1}$: accounting for $\mathsf{L}$ leads to subleading corrections in $k$.     Inverting the remaining $6\times 6$ matrix and keeping only the leading order terms as $k\rightarrow 0$, we obtain
\begin{equation}
 \mathcal{A}_{ij}(k) = \frac{2e^2v_{\mathrm{F,2}}\left(v_{\mathrm{F,1}}^2 + v_{\mathrm{F,2}}^2\right)^2}{\ell_{\mathrm{ee}} \left(v_{\mathrm{F,1}}^4 + v_{\mathrm{F,2}}^4\right)k^2} \left(\mdelta_{ij} - \frac{k_ik_j}{k^2}\right),  \label{eq:appd}
 \end{equation}
 which has the expected $k^{-2}$ dependence arising from transverse momentum diffusion.   Using the generalized techniques of \cite{levitov1607, lucas1612, hartnoll1706}, we have numerically evaluated $\mathcal{A}_{ij}$ for any value of $k\ell_{\mathrm{ee}}$.   This result is plotted in Figure \ref{fig:res} of the main text.

\section{Normal  Modes When $b=0$}\label{app:b0}
In this appendix, we explicitly give the approximate eigenvalues and eigenvectors of $\mathsf{W}+\mathsf{L}(\mathbf{k})$, corresponding to the gapless hydrodynamic modes.   The techniques to derive this result are analogous to those used to derive (\ref{eq:appd});  for this model, the ``top block" of (\ref{eq:blockmatrix}) consists of modes with $|j| \le 2$.   Without loss  of generality, we take $k_x=k$, $k_y=0$.  In the hydrodynamic limit $k \ell_{\mathrm{ee}} \rightarrow 0$,  we find  \begin{subequations}\begin{align}
\omega &=  \pm \frac{v_{\mathrm{F}}}{2}k + \cdots, \;\;\;\;\;  \text{eigenvector} = |2\rangle \pm |1\rangle  \mp |-1\rangle - |-2\rangle, \\
\omega &= \pm \frac{\sqrt{3}v_{\mathrm{F}}}{2} k + \cdots, \;\;\;\;\;  \text{eigenvector} = |2\rangle \pm \sqrt{3} |1\rangle  + 2|0\rangle \pm  \sqrt{3}  |-1\rangle + |-2\rangle, \\
\omega &= - \mathrm{i} \frac{v_{\mathrm{F}}^2}{6\gamma} k^2 + \cdots, \;\;\;\;\;  \text{eigenvector} = |2\rangle - |0\rangle + |-2\rangle.
\end{align}\end{subequations}
Note that there are no hydrodynamic modes involving $|j\rangle$ for $|j|>2$ -- all of these modes have a finite lifetime $\omega \approx -\mathrm{i}v_{\mathrm{F}}/\ell_{\mathrm{ee}}$.    We observe  that  the current  modes $|\pm 1\rangle$ are  only  included in ballistically propgating modes (generalized ``sound waves").   So, as asserted  in the main text, random magnetic fields  do  not excite diffusive fluctuations, and this is why $\rho \propto \ell_{\mathrm{ee}}$ in this  model.   In contrast, chemical potential disorder couples to $|0\rangle$, not $|\pm 1\rangle$.   $|0\rangle$ does have overlap with  the diffusive mode above, and this is why (as shown in Figure \ref{fig:res}), potential disorder leads to (\ref{eq:main}) in  this model.
\end{appendix}

\bibliographystyle{unsrt}
\addcontentsline{toc}{section}{References}
\bibliography{kineticmagbib}

\begin{thebibliography}{10}

\bibitem{kadowaki}
K.~Kadowaki and S.~B. Woods.
\newblock ``Universal relationship of the resistivity and specific heat in
  heavy-fermion compounds",
  \href{https://doi.org/10.1016/0038-1098(86)90785-4}{\textsl{Solid State
  Communications} \textbf{58} 507 (1987)}.

\bibitem{jacko}
A.~C. Jacko, J.~O. Fjaerestad, and B.~J. Powell.
\newblock ``Unified explanation of the Kadowaki-Woods ratio in strongly
  correlated materials",
  \href{https://doi.org/10.1038/nphys1249}{\textsl{Nature Physics} \textbf{5}
  422 (2009)}, \href{http://arxiv.org/abs/0805.4275}{\texttt{arXiv:0805.4275}}.

\bibitem{mackenzie2013}
J.~A.~N. Bruin, H.~Sakai, R.~S. Perry, and A.~P. Mackenzie.
\newblock ``Similarity of scattering rates in metals showing $T$-linear
  resistivity",
  \href{http://science.sciencemag.org/content/339/6121/804}{\textsl{Science}
  \textbf{339} 804 (2013)}.

\bibitem{hartnoll1}
S.~A. Hartnoll.
\newblock ``Theory of universal incoherent metallic transport",
  \href{http://www.nature.com/nphys/journal/v11/n1/full/nphys3174.html}{\textsl{Nature
  Physics} \textbf{11} 54 (2015)},
  \href{http://arxiv.org/abs/1405.3651}{\texttt{arXiv:1405.3651}}.

\bibitem{ziman}
J.~Ziman.
\newblock \emph{Electrons and Phonons}
  \href{http://www.amazon.com/Electrons-Phonons-Transport-Phenomena-Physical/dp/0198507798/ref=sr_1_1?ie=UTF8&qid=1433301493&sr=8-1&keywords=electrons+and+phonons&pebp=1433301494471&perid=0S820PYT3GQQ1X5Q7Z92}{(Oxford
  University Press, 1960)}.

\bibitem{baber}
W.~G. Baber.
\newblock ``The contribution to the electrical resistance of metals from
  collisions between electrons",
  \href{https://doi.org/10.1098/rspa.1937.0027}{\textsl{Proceedings of the
  Royal Society} \textbf{A158} 383 (1937)}.

\bibitem{behnia}
X.~Lin, B.~Fauque, and K.~Behnia.
\newblock ``Scalable $T^2$ resistivity in a small single-component Fermi
  surface", \href{https://doi.org/10.1126/science.aaa8655}{\textsl{Science}
  \textbf{349} 945 (2015)},
  \href{http://arxiv.org/abs/1508.07812}{\texttt{arXiv:1508.07812}}.

\bibitem{stemmer}
E.~Mikheev, S.~Raghavan, J.~Y. Zhang, P.~B. Marshall, A.~P. Kajdos, L.~Balents,
  and S.~Stemmer.
\newblock ``Carrier density independent scattering rate in
  $\mathrm{SrTiO}_3$-based electron liquids",
  \href{https://doi.org/10.1038/srep20865}{\textsl{Scientific Reports}
  \textbf{6} \texttt{20865} (2016)},
  \href{http://arxiv.org/abs/1512.02294}{\texttt{arXiv:1512.02294}}.

\bibitem{gurzhi}
R.~N. Gurzhi.
\newblock ``Minimum of resistance in impurity-free conductors",
  \href{http://www.jetp.ac.ru/cgi-bin/e/index/e/17/2/p521?a=list}{\textsl{Journal
  of Experimental and Theoretical Physics} \textbf{17} 521 (1963)}.

\bibitem{spivak02}
M.~Hruska and B.~Spivak.
\newblock ``Conductivity of the classical two-dimensional electron gas",
  \href{http://journals.aps.org/prb/abstract/10.1103/PhysRevB.65.033315}{\textsl{Physical
  Review} \textbf{B65} \texttt{033315} (2002)},
  \href{http://arxiv.org/abs/cond-mat/0102219}{\texttt{arXiv:cond-mat/0102219}}.

\bibitem{andreev}
A.~V. Andreev, S.~A. Kivelson, and B.~Spivak.
\newblock ``Hydrodynamic description of transport in strongly correlated
  electron systems",
  \href{http://journals.aps.org/prl/abstract/10.1103/PhysRevLett.106.256804}{\textsl{Physical
  Review Letters} \textbf{106} \texttt{256804} (2011)},
  \href{http://arxiv.org/abs/1011.3068}{\texttt{arXiv:1011.3068}}.

\bibitem{lucas}
A.~Lucas.
\newblock ``Hydrodynamic transport in strongly coupled disordered quantum field
  theories",
  \href{http://iopscience.iop.org/article/10.1088/1367-2630/17/11/113007/meta}{\textsl{New
  Journal of Physics} \textbf{17} \texttt{113007} (2015)},
  \href{http://arxiv.org/abs/1506.02662}{\texttt{arXiv:1506.02662}}.

\bibitem{polini}
I.~Torre, A.~Tomadin, A.~K. Geim, and M.~Polini.
\newblock ``Non-local transport and the hydrodynamic shear viscosity in
  graphene",
  \href{http://journals.aps.org/prb/abstract/10.1103/PhysRevB.92.165433}{\textsl{Physical
  Review} \textbf{B92} \texttt{165433} (2016)},
  \href{http://arxiv.org/abs/1508.00363}{\texttt{arXiv:1508.00363}}.

\bibitem{levitovhydro}
L.~Levitov and G.~Falkovich.
\newblock ``Electron viscosity, current vortices and negative nonlocal
  resistance in graphene",
  \href{http://www.nature.com/nphys/journal/v12/n7/full/nphys3667.html}{\textsl{Nature
  Physics} \textbf{12} 672 (2016)},
  \href{http://arxiv.org/abs/1508.00836}{\texttt{arXiv:1508.00836}}.

\bibitem{lucas3}
A.~Lucas, J.~Crossno, K.~C. Fong, P.~Kim, and S.~Sachdev.
\newblock ``Transport in inhomogeneous quantum critical fluids and in the Dirac
  fluid in graphene",
  \href{http://journals.aps.org/prb/abstract/10.1103/PhysRevB.93.075426}{\textsl{Physical
  Review} \textbf{B93} \texttt{075426} (2016)},
  \href{http://arxiv.org/abs/1510.01738}{\texttt{arXiv:1510.01738}}.

\bibitem{alekseev}
P.~S. Alekseev.
\newblock ``Negative magnetoresistance in viscous flow of two-dimensional
  electrons",
  \href{http://journals.aps.org/prl/abstract/10.1103/PhysRevLett.117.166601}{\textsl{Physical
  Review Letters} \textbf{117} \texttt{166601} (2016)}.

\bibitem{levitov1607}
H.~Guo, E.~Ilseven, G.~Falkovich, and L.~Levitov.
\newblock ``Higher-than-ballistic conduction of viscous electron flows",
  \href{http://www.pnas.org/content/114/12/3068.abstract}{\textsl{Proceedings
  of the National Academy of Sciences} \textbf{114} 3068 (2017)},
  \href{http://arxiv.org/abs/1607.07269}{\texttt{arXiv:1607.07269}}.

\bibitem{lucas1612}
A.~Lucas.
\newblock ``Stokes paradox in electronic Fermi liquids",
  \href{http://link.aps.org/doi/10.1103/PhysRevB.95.115425}{\textsl{Physical
  Review} \textbf{B95} \texttt{115425} (2017)},
  \href{http://arxiv.org/abs/1612.00856}{\texttt{arXiv:1612.00856}}.

\bibitem{levitov1612}
H.~Guo, E.~Ilseven, G.~Falkovich, and L.~Levitov.
\newblock ``Stokes paradox, back reflections and interaction-enhanced
  conductance",
  \href{http://arxiv.org/abs/1612.09239}{\texttt{arXiv:1612.09239}}.

\bibitem{hartnoll1704}
A.~Lucas and S.~A. Hartnoll.
\newblock ``Resistivity bound for hydrodynamic bad metals",
  \href{http://arxiv.org/abs/1704.07384}{\texttt{arXiv:1704.07384}}.

\bibitem{hartnoll1706}
A.~Lucas and S.~A. Hartnoll.
\newblock ``Kinetic theory of transport for inhomogeneous electron fluids",
  \href{http://arxiv.org/abs/1706.04621}{\texttt{arXiv:1706.04621}}.

\bibitem{lucasreview2}
A.~Lucas and K.~C. Fong.
\newblock ``Hydrodynamics of electrons in graphene",
  \href{http://arxiv.org/abs/1710.08425}{\texttt{arXiv:1710.08425}}.

\bibitem{molenkamp}
M.~J.~M. de~Jong and L.~W. Molenkamp.
\newblock ``Hydrodynamic electron flow in high-mobility wires",
  \href{http://journals.aps.org/prb/abstract/10.1103/PhysRevB.51.13389}{\textsl{Physical
  Review} \textbf{B51} 11389 (1995)},
  \href{http://arxiv.org/abs/cond-mat/9411067}{\texttt{arXiv:cond-mat/9411067}}.

\bibitem{bandurin}
D.~A.~Bandurin \emph{et al.}
\newblock ``Negative local resistance due to viscous electron backflow in
  graphene",
  \href{http://science.sciencemag.org/content/351/6277/1055}{\textsl{Science}
  \textbf{351} 1055 (2016)},
  \href{http://arxiv.org/abs/1509.04165}{\texttt{arXiv:1509.04165}}.

\bibitem{crossno}
J.~Crossno \emph{et al.}
\newblock ``Observation of the Dirac fluid and the breakdown of the
  Wiedemann-Franz law in graphene",
  \href{http://science.sciencemag.org/content/351/6277/1058}{\textsl{Science}
  \textbf{351} 1058 (2016)},
  \href{http://arxiv.org/abs/1509.04713}{\texttt{arXiv:1509.04713}}.

\bibitem{mackenzie}
P.~J.~W. Moll, P.~Kushwaha, N.~Nandi, B.~Schmidt, and A.~P. Mackenzie.
\newblock ``Evidence for hydrodynamic electron flow in $\mathrm{PdCoO}_2$",
  \href{http://science.sciencemag.org/content/351/6277/1061}{\textsl{Science}
  \textbf{351} 1061 (2016)},
  \href{http://arxiv.org/abs/1509.05691}{\texttt{arXiv:1509.05691}}.

\bibitem{felser}
J.~Gooth, F.~Menges, C.~Shekhar, V.~S\"uss, N.~Kumar, Y.~Sun, U.~Drechsler,
  R.~Zierold, C.~Felser, and B.~Gotsmann.
\newblock ``Electrical and thermal transport at the Planckian bound of
  dissipation in the hydrodynamic electron fluid of $\mathrm{WP}_2$",
  \href{http://arxiv.org/abs/1706.05925}{\texttt{arXiv:1706.05925}}.

\bibitem{novikov}
D.~S. Novikov.
\newblock ``Viscosity of a two-dimensional Fermi liquid",
  \href{http://arxiv.org/abs/cond-mat/0603184}{\texttt{arXiv:cond-mat/0603184}}.

\bibitem{levitov1703}
R.~Krishna~Kumar \emph{et al}.
\newblock ``Super-ballistic flow of viscous electron fluid through graphene
  constrictions",
  \href{http://arxiv.org/abs/1703.06672}{\texttt{arXiv:1703.06672}}.

\bibitem{kwwest}
Q.~Shi, P.~D. Martin, Q.~A. Ebner, M.~A. Zudov, L.~N. Pfeiffer, and K.~W. West.
\newblock ``Colossal negative magnetoresistance in a two-dimensional electron
  gas",
  \href{http://journals.aps.org/prl/abstract/10.1103/PhysRevB.89.201301}{\textsl{Physical
  Review} \textbf{B89} \texttt{201301} (2014)}.

\bibitem{doniach}
S.~Doniach.
\newblock ``The Kondo lattice and weak antiferromagnetism",
  \href{https://doi.org/10.1016/0378-4363(77)90190-5}{\textsl{Physica}
  \textbf{B91} 231 (1977)}.

\bibitem{aynajian}
P.~Aynajian, E.~H. da~Silva~Neto, A.~Gyenis, R.~E. Baumbach, J.~D. Thompson,
  Z.~Fisk, E.~D. Bauer, and A.~Yazdani.
\newblock ``Visualizing heavy fermions emerging in a quantum critical Kondo
  lattice", \href{https://doi.org/10.1038/nature11204}{\textsl{Nature}
  \textbf{486} 201 (2012)},
  \href{http://arxiv.org/abs/1206.3145}{\texttt{arXiv:1206.3145}}.

\bibitem{ypwu}
Y.~P. Wu, D.~Zhao, A.~F. Wang, N.~Z. Wang, Z.~J. Xiang, X.~G. Luo, T.~Wu, and
  X.~H. Chen.
\newblock ``Emergent Kondo lattice behavior in iron-based superconductors
  $\mathrm{AFe}_2\mathrm{As}_2$ ($\mathrm{A}=\mathrm{K}$, Rb, Cs)",
  \href{https://doi.org/PhysRevLett.116.147001}{\textsl{Physical Review
  Letters} \textbf{116} \texttt{147001} (2016)},
  \href{http://arxiv.org/abs/1507.08732}{\texttt{arXiv:1507.08732}}.

\bibitem{alloul}
H.~Alloul, P.~Mendels, H.~Casalta, J.~F. Marucco, and J.~Arabski.
\newblock ``Correlations between magnetic and superconducting properties of
  Zn-substituted $\mathrm{YBa}_2\mathrm{Cu}_3\mathrm{O}_{6+x}$",
  \href{https://doi.org/10.1103/PhysRevLett.67.3140}{\textsl{Physical Review
  Letters} \textbf{67} 3140 (1991)}.

\bibitem{mendels}
P.~Mendels, J.~Bobroff, G.~Collin, H.~Alloul, M.~Gabay, J.~F. Marucco,
  N.~Blanchard, and B.~Greiner.
\newblock ``Normal-state magnetic properties of Ni and Zn substituted in
  $\mathrm{YBa}_2\mathrm{Cu}_3\mathrm{O}_{6+x}$: hole-doping dependence",
  \href{https://doi.org/10.1209/epl/i1999-00319-x}{\textsl{Europhysics Letters}
  \textbf{46} 678 (1999)}.

\bibitem{balatsky}
A.~V. Balatsky, I.~Vekhter, and J-X. Zhu.
\newblock ``Impurity-induced states in conventional and unconventional
  superconductors",
  \href{https://doi.org/10.1103/RevModPhys.78.373}{\textsl{Reviews of Modern
  Physics} \textbf{78} 373 (2006)},
  \href{http://arxiv.org/abs/cond-mat/0411318}{\texttt{arXiv:cond-mat/0411318}}.

\bibitem{brinkman}
A.~Brinkman, M.~Huijben, M.~van Zalk, J.~Huijben, U.~Zeitler, J.~C. Maan, W.~G.
  van~der Wiel, G.~Rijnders, D.~H.~A. Blank, and H.~Hilgenkamp.
\newblock ``Magnetic effects at the interface between nonmagnetic oxides",
  \href{https://doi.org/10.1038/nmat1931}{\textsl{Nature Materials} \textbf{6}
  493 (2007)},
  \href{http://arxiv.org/abs/cond-mat/0703028}{\texttt{arXiv:cond-mat/0703028}}.

\bibitem{menyoung}
M.~Lee, J.~R. Williams, S.~Zhang, C.~D. Frisbie, and D.~Goldhaber-Gordon.
\newblock ``Electrolyte gate-controlled Kondo effect in $\mathrm{SrTiO}_3$",
  \href{https://doi.org/10.1103/PhysRevLett.107.256601}{\textsl{Physical Review
  Letters} \textbf{107} \texttt{256601} (2011)},
  \href{http://arxiv.org/abs/1108.0139}{\texttt{arXiv:1108.0139}}.

\bibitem{moler}
B.~Kalisky, J.~A. Bert, C.~Bell, Y.~Xie, H.~K. Sato, M.~Hosoda, Y.~Hikita,
  H.~Y. Hwang, and K.~A. Moler.
\newblock ``Scanning probe manipulation of magnetism at the
  $\mathrm{LaAlO}_3/\mathrm{SrTiO}_3$ heterointerface",
  \href{https://doi.org/10.1021/nl301451e}{\textsl{Nano Letters} \textbf{12}
  4055 (2012)}.

\bibitem{kamenev}
A.~Kamenev.
\newblock \emph{Field Theory of Non-Equilibrium Systems}
  \href{https://www.amazon.com/Field-Theory-Non-Equilibrium-Systems-Kamenev/dp/0521760828/ref=sr_1_1?ie=UTF8&qid=1478548419&sr=8-1&keywords=field+theory+of+non-equilibrium+systems}{(Cambridge
  University Press, 2011)}.

\bibitem{hofman}
S.~A. Hartnoll and D.~M. Hofman.
\newblock ``Locally critical umklapp scattering and holography",
  \href{https://doi.org/10.1103/PhysRevLett.108.241601}{\textsl{Physical Review
  Letters} \textbf{108} \texttt{241601} (2012)},
  \href{http://arxiv.org/abs/1201.3917}{\texttt{arXiv:1201.3917}}.

\bibitem{lucasMM}
A.~Lucas and S.~Sachdev.
\newblock ``Memory matrix theory of magnetotransport in strange metals",
  \href{http://journals.aps.org/prb/abstract/10.1103/PhysRevB.91.195122}{\textsl{Physical
  Review} \textbf{B91} \texttt{195122} (2015)},
  \href{http://arxiv.org/abs/1502.04704}{\texttt{arXiv:1502.04704}}.

\bibitem{lucasreview}
S.~A. Hartnoll, A.~Lucas, and S.~Sachdev.
\newblock ``Holographic quantum matter",
  \href{http://arxiv.org/abs/1612.07324}{\texttt{arXiv:1612.07324}}.

\bibitem{bgk}
P.~L. Bhatnagar, E.~P. Gross, and M.~Krook.
\newblock ``A model for collision processes in gases. I. Small amplitude
  processes in charged and neutral one-component systems",
  \href{https://journals.aps.org/pr/abstract/10.1103/PhysRev.94.511}{\textsl{Physical
  Review} \textbf{94} 511 (1954)}.

\bibitem{ledwith1}
P.~Ledwith, H.~Guo, and L.~Levitov.
\newblock ``Fermion collisions in two dimensions",
  \href{http://arxiv.org/abs/1708.01915}{\texttt{arXiv:1708.01915}}.

\bibitem{ledwith2}
P.~Ledwith, H.~Guo, A.~V. Shytov, and L.~Levitov.
\newblock ``Head-on collisions and scale-dependent viscosity in two-dimensional
  electron systems",
  \href{http://arxiv.org/abs/1708.02376}{\texttt{arXiv:1708.02376}}.

\bibitem{kadanoff}
L.~P. Kadanoff and P.~C. Martin.
\newblock ``Hydrodynamic equations and correlation functions",
  \href{http://www.sciencedirect.com/science/article/pii/0003491663900782}{\textsl{Annals
  of Physics} \textbf{24} 419 (1963)}.

\bibitem{kovtun}
P.~Kovtun.
\newblock ``Lectures on hydrodynamic fluctuations in relativistic theories",
  \href{http://m.iopscience.iop.org/1751-8121/45/47/473001/}{\textsl{Journal of
  Physics} \textbf{A45} \texttt{473001} (2012)},
  \href{http://arxiv.org/abs/1205.5040}{\texttt{arXiv:1205.5040}}.

\bibitem{dresselhaus}
M.~S. Dresselhaus, G.~Dresselhaus, and A.~Jorio.
\newblock \emph{Group Theory: Application to the Physics of Condensed Matter},
  \href{https://www.amazon.com/Group-Theory-Application-Physics-Condensed/dp/3642069452}{(Springer,
  2010)}.

\end{thebibliography}

\end{document}